\documentclass{elsart}

 \usepackage{harvard}

\usepackage{amssymb}

\def\url#1{{\ttfamily\def\/{/\discretionary{}{}{}}#1}}

\begin{document}

\begin{frontmatter}
\title{Astrocomp: a web service for the use of high performance
computers in Astrophysics}

\author[ct]{U. Becciani},
\author[uniroma]{R. Capuzzo Dolcetta},
\author[ct]{A. Costa},
\author[uniroma]{P. Di Matteo},
\author[uniroma]{P. Miocchi},
\author[enea]{V. Rosato}

\address[ct]{INAF-Osservatorio Astrofisico di Catania, Via Santa Sofia 78,\\
95123 Catania, Italy}
\address[uniroma]{Dipartimento di Fisica, Universit\'a di Roma ``La Sapienza'', P.le Aldo Moro 2,\\ 00185 Roma, Italy}
\address[enea]{ENEA Casaccia Research Center, Via Anguillarese 301, Roma, Italy}
\begin{abstract}
Astrocomp is a joint project, developed by the INAF-Astrophysical Observatory of Catania, University of Roma La Sapienza and Enea.
The project has the goal of providing the scientific community of
a web-based user-friendly interface which
allows running parallel codes on a set of high-performance computing (HPC) resources, without any need for
specific knowledge about parallel programming and Operating Systems commands.
Astrocomp provides, also, computing time on a set of  parallel computing systems, available to the authorized user.
At present, the portal makes a few codes available, among which: FLY, a cosmological code for studying
three-dimensional collisionless self-gravitating systems with periodic boundary conditions; ATD, a parallel
tree-code  for the simulation of the dynamics of boundary-free collisional and collisionless
self-gravitating systems and MARA, a code for stellar light curves analysis.
Other codes are going to be added to the portal.

\end{abstract}

\end{frontmatter}


\section{Introduction}
Many  codes have been produced by the astronomical community, which address many different
aspects related to the modelling of astrophysical phenomena. Most of the time these codes are
built as software packages, which can then be downloaded from the interested user on his/her
own computing system and installed. Sometimes these packages are packed as precompiled
``distributions'', for specific architectures and/or operating systems, containing
already binaries and executable files which, at the end of the installation process,
are put in some convenient areas. In other cases, the source codes are given,
and instructions for compiling them on different systems are provided. Anyhow,
the final user is asked to install the software directly on a computer system available.\\
On the other hand, the diffusion of HTML languages and of the related technologies have
made feasible a different approach to code usage, where one does not need anymore
to download and install locally large software packages. This approach makes use
of {\em portals}, i.e. some multifunction WWW interfaces, to allow remote users to
prepare and execute some computational tasks on some predetermined platforms,
where the user has been authorised to execute his/her jobs. The advantage is that
the user is relieved from the duty of downloading and compiling the source code:
the latter phase can sometimes be rather cumbersome, particularly to an
unexperienced user. The obvious disadvantage of this approach is that the user has
lesser control over the code he is using, particularly because he can operate
very few (if any) changes on the source code.\\
Usually, Web portals for the astronomical community have been mostly designed and
used to enable database searches. A complex extension of this concept, namely the
Astrophysical Virtual Observatory (AVO, http://www.euro-vo.org/) is currently
under active development as a joint, large-scale effort involving few international
Institutions. The project presente here, Astrocomp (http://www.astrocomp.it),
is a portal specifically designed
to allow the access to {\it simulation} codes, rather than pre-existing databases.
The end user will be able to remotely prepare the simulation, after browsing
among those existing on the specified platforms, perform the run, and finally
retrieve the data, interactively operating with the portal.\\
This paper introduces the portal, its main functionalities and  gives a short
description of the software presently available (more codes can be added on
request of the authors). Section 2 contains a description of the portal, while in
Sect. 3 an account of the languages and technologies used to build the portal is present.
Sect. 4 describes characteristics of the numerical codes available, while Sect. 5
gives introductory info on the practical use of the facilities of the portal.
Finally, Section 6 contains the conclusions.

\section{The Astrocomp portal}
The Astrocomp project is being developed by the INAF-Astrophysical Observatory of Catania,
the University of Roma La Sapienza
 and ENEA,  funded by a Consiglio Nazionale delle Ricerche (CNR)
grant  in the frame of  the program `Agenzia 2000'.
\\ Astrocomp is a portal  (http://www.astrocomp.it) based on WEB  technologies, aimed at managing
and using codes for Astrophysics, employing user-friendly interfaces; at present,
the codes available are, mainly, related to the simulation of gravitational systems of
astrophysical and cosmological interest.
The main functionality of the portal  is a user-friendly application which allows
the scientist to submit a job  in a grid of computing systems. Indeed,
even if, at the moment, there is a direct link between the software and the system
where the user job will run, the database architecture is not related to a specific
system platform, and the connection with the system is made with several parameters that can be easily changed.
\\ At present, the computational platforms registered in  Astrocomp are the
CINECA (Casalecchio di Reno, Bologna, Italy; http://www.cineca.it)  MPP systems:
IBM SP4 with 512 PEs and IBM Linux cluster with 512 PEs
through the formal agreement between INAF (Istituto Nazionale di Astrofisica)
and CINECA, and the IBM SP system with 32 PEs sited at the Catania Astrophysical Observatory. Registered Astrocomp
users can freely use them (within the assigned time quota).
\section{The portal structure and resources}
Astrocomp runs on  Apache-Advanced Extranet\footnote{ ADVX, see 
http://www.advx.org/}: this
allows to set up connections with no
practical limits on the number of authorized users.
\\ Apache is a powerful, flexible,
compliant web server, it has a flexible configuration and it is extensible to
third-party modules. The access to the Astroadmin section, i.e. the reserved area written for the Administrator for  Authorization and Authentication,
is based on the Apache facilities.
Dynamical web pages are realized with a server-side scripting PHP (recursive acronym for
``PHP: Hypertext Preprocessor''), a widely used Open Source general-purpose scripting language
that is especially suited for Web development and can be embedded in HTML.
The PHP programming language provides a powerful extension to HTML to create advanced
and interactive web pages.
\\ Astrocomp has a database of the software and of the  systems and codes managed by the portal;
this is done with the MySQL database management system.
The database tables contain a description of the codes properties and of the hardware resources.
The portal is designed with a relational database containing the information related
to the hardware systems accessible to Astrocomp, that is arranged such to
ensure an easy implementation of new HPC resources. The Astrocomp system
administrator updates the hardware table and in particular fills the fields related to OS commands:
user login, directories, disk quota, job submission and job management etc.

The administration of all the portal features is done through a reserved section.
A new code can be quickly added to the  portal and its behaviour can be controlled
on the user-side.
There are no practical limits to the variety of codes that can be included in the
Astrocomp structure.\\
The portal architecture allows us to handle each code
 considering the I/O files, shell scripts, boundary condition data, log files, etc.
A job is an entity described by a state: Idle, Queued or Running and some other parameters.
An Astrocomp section allows the user to know the system status of an HPC giving updated
information like memory and cpu usage.
\section{Software Resources}
%
In the following Sections we shortly describe the codes that are presently available in Astrocomp.
More exhaustive descriptions of these codes can be found in the portal.
%
\subsection{FLY}
FLY is a tree parallel code that follows the evolution of
a newtonian three-dimensional N-body collisionless system.
It is based on the tree Barnes-Hut algorithm (\citeasnoun{BH}) and periodical boundary conditions are
implemented by means of Ewald's (\citeasnoun{Ewa})  summation technique.
The evolutive differential equations are integrated using the classical (second order)
`leap-frog' integration scheme with a fixed time-step.
All the particles have assigned the same mass, and the spatial cubic grid
has fixed size.
The supported cosmological models are: Standard Cold Dark Matter
 $\Omega_{TOT}=1$,
Lambda Models (with Cold Dark Matter), and Open Models $\Omega_{TOT}<1$.
\\ The code was originally developed on a CRAY T3E system using
the logically SHared MEMory access routines (SHMEM)
but it also runs on SGI ORIGIN systems and on IBM SP by using Low-Level Application
 Programming Interface routines (LAPI).
Fly is included in the CPC (Computer Physics Communications) program Library.
More details can be found in  \citeasnoun{ube} and  at http://www.ct.astro.it/fly/.

%
%

\subsection{ATD}
ATD (Adaptive Tree Decomposition tree-code) is a parallel N-body code
for the simulations of the dynamics of collisonal and collisionless boundary-free self-gravitating stellar systems.
The  gravitational interaction among the `particles' is computed
with an algorithm based on the Barnes \& Hut scheme (\citeasnoun{BH})
including up to the quadrupole
moment in the multipolar expansion.
A `dynamical' tree reconstruction is also implemented in the code. This permits a more
frequent update of the small--scale (neighbouring) interactions
(evaluated by direct summation) in respect to the large--scale forces, without the need
of a complete reconstruction of the whole data tree--structure.
The time integration of the trajectories of the particles representing
stars in the numerical model, is performed using individual and variable time-steps
in a  `leap-frog' algorithm  corrected in such a way to preserve 2$^{nd}$ order
accuracy also during the time-step change.

The code has been parallelized adopting an original scheme for the distribution of
the computational work among processors called `Adaptivetree decomposition' (interested
readers can find an exhaustive description of this scheme in \citeasnoun{MCD}).
We carried out two different parallel versions of the code which can run both on
shared memomy platforms (using OpenMP language directives) and on distributed
memory computers (employing suitable MPI calls).
Further details on the code features and its usage can be found
in the portal itself.
\subsection{MARA}
MARA is an MPI code for modelling sequences of light curves of single and binary stars with surface
brightness inhomogeneities, in particular cool spots. It allows  to derive sequences of maps of the spot
surface distribution which are useful to study magnetic solar-like activity in close binaries and single stars. Moreover,
it allows  to determine the best set of photometric parameters for a close active binary
by correcting the systematic errors due to the light curve distortion induced by spots
(\citeasnoun{RLB}; \citeasnoun{LB}).
\section{An introduction to the use of Astrocomp}
%

Anyone  interested in using the facilities provided by Astrocomp has to
register, first. The steps to do  are:
\begin{enumerate}
\item register as a new user by clicking on the {\tt Registration Form} button on
the left-hand side of the home-page;
\item fill in carefully the form and submit it;
\item wait for a confirmation e-mail sent by the portal administrator.
\end{enumerate}

The Scientific staff will evaluate the request and assign the access to the portal
services according to three different `user classes': A,B and C.
The A class  has the highest limit on the CPU time and disk usage;
the B class supplies the user with a limited amount of CPU time and disk quota. while the C class
assigns only few resources to execute very short jobs.

Clicking on the {\tt User Area} button, any registered user can have the
access to all the computational facilities of AstroComp, after the normal
 login procedure.
At present, the AstroComp user is allowed to:
\begin{itemize}
\item start a new simulation by choosing the code  to  used and by
determining the simulation parameters via the {\tt Parameters} on-line form
(initial conditions  can also be uploaded by the local user's host to the server);
\item choose among different platforms in the pool of the available
resources of AstroComp,
taking into account the work-load and the
accessibility of each system;
\end{itemize}
or
\begin{itemize}
\item browse the status of a previously launched job, possibly checking
the intermediate results with a preliminary `on-the-fly' visualization
tool already available in the portal; 
\item download the final and/or intermediate results.
\end{itemize}

Moreover, anyone who visits the portal without being, necessarily, a registered user
can also examine all the implemented software
clicking on the {\tt Software} button on the home-page (http://www.astrocomp.it/software),
read the enclosed documentations and manuals, and take a look at the features and the status
of all the computing machines that can be employed clicking on {\tt Hardware}.
(http://www.astrocomp.it/user/hardware)

%
%


%
%
%
%

%
The user chooses the system where the job will be submitted, then he fills in some
forms specifying the parameters and all the variables involved in the job.
The complete job history is stored in a MySQL table.
The user can also easily retrieve the parameter collection and can re-use them for
a new job submission. \\ 
The last phase consists in the job preparation and submission 
%

It is the portal itself that copies and/or uploads to the remote system, all the needed files
and shell scripts and compiles and submits the code.
In the next future the grid environement will allow the user to register and store the
output files in a Storage Element (e.g. DataGrid environment, see http://eu-datagrid.web.cern.ch/).


%

\section{Conclusions}
The aim of Astrocomp (http://www.astrocomp.it) is to run Astrophysical codes on a grid of systems,
avoiding the effort to learn new specific Operating System and/or parallelization commands.
Even if, at  present, Astrocomp allows  the execution on a limited set
of systems,
according to the design of the portal and its future development we foresee
to execute Astrocomp
on an effective computational grid. CINECA (http://www.cineca.it) has already selected this
project as a pilot project for
internal MPP grid usage. A formal agreement between the Italian National Institute of Nuclear Physics (INFN)
and the INAF-Astrophysical Observatory of Catania allows us to build a local DataGrid node at the INAF site.
Astrocomp will also be ported to the DataGrid in the next future.
\\ All scientists interested in the Astrocomp facility are welcome and  invited  to ask for the inclusion of
their own software in the Astrocomp database.
%

\end{document}